\documentclass{svmult}

\usepackage{graphicx}
\usepackage{makeidx}
\usepackage{multicol}
\usepackage{footmisc}
\usepackage{amssymb}
\usepackage[twoside,closeFloats]{fltpage}

\begin{document}

\title*{Common Envelope Evolution Redux}
\author{Ronald F. Webbink}
\institute{Department of Astronomy, University of Illinois, 1002 W. Green St.,
Urbana, IL  61801, USA}
\maketitle

\begin{abstract}
Common envelopes form in dynamical time scale mass exchange, when the envelope
of a donor star engulfs a much denser companion, and the core of the donor
plus the dense companion star spiral inward through this dissipative envelope. 
As conceived by Paczynski and Ostriker, this process must be responsible for
the creation of short-period binaries with degenerate components, and, indeed,
it has proven capable of accounting for short-period binaries containing one
white dwarf component.  However, attempts to reconstruct the evolutionary
histories of close double white dwarfs have proven more problematic, and point
to the need for enhanced systemic mass loss, either during the close of the
first, slow episode of mass transfer that produced the first white dwarf, or
during the detached phase preceding the final, common envelope episode.  The
survival of long-period interacting binaries with massive white dwarfs, such
as the recurrent novae T CrB and RS Oph, also presents interpretative
difficulties for simple energetic treatments of common envelope evolution. 
Their existence implies that major terms are missing from usual formulations
of the energy budget for common envelope evolution.  The most plausible
missing energy term is the energy released by recombination in the common
envelope, and, indeed, a simple reformulation the energy budget explicitly
including recombination resolves this issue.
\end{abstract}

\section{Introduction}
\label{Intro}

From the realization~\cite[et seq.]{Kra62,Kra64} that all cataclysmic
variables (CVs) are interacting binary stars, their existence posed a dilemma
for theories of binary evolution.  The notion that close binary stars might
evolve in ways fundamentally different from isolated stars was rooted in the
famous `Algol paradox' (that the cooler, lobe-filling subgiant or giant
components among these well-known eclipsing binaries are less massive, but
more highly evolved, than their hotter main-sequence companions).  The
resolution of that paradox invoked large-scale mass transfer reversing the
initial mass ratios of these binaries~\cite{Mor60}.  Indeed, model
calculations assuming conservation of total mass and orbital angular momenum
are qualitatively consistent with the main features of Algol-type binaries. 
Even if quantitative consistency between models and observational data
generally requires some losses of mass and angular momentum among Algol
binaries (e.g.,~\cite{Giu81,Ibe84,Egg06}), the degree of those losses is
typically modest, and the remnant binary is expected to adhere closely to an
equilibrium core mass-radius relation for low-mass giant stars (see, e.g., the
pioneering study of AS Eri by Refsdal, Roth \& Weigert~\cite{Ref74}).  Those
remnant binaries are typically of long orbital period (days to weeks) in
comparison with CVs, and furthermore typically contain helium white dwarfs of
low mass, especially in the short-period limit.  In contrast, CVs evidently
contain relatively massive white dwarfs, in binary systems of much shorter
orbital periods (hours), that is, with much smaller total energies and orbital
angular momenta.

In an influential analysis of the Hyades eclipsing red dwarf/white dwarf
binary BD +16$^\circ$ 516 (= V471 Tau), Vauclair~\cite{Vau72} derived a total
system mass less than the turnoff mass of the Hyades, and noted that the
cooling age of the white dwarf component was much smaller than the age of the
cluster.  He speculated that V471 Tau in its present state was the recent
product of the ejection of a planetary nebula by the white dwarf. 
Paczynski~\cite{Pac76} realized that, immediately prior to that event, the
white dwarf progenitor must have been an asymptotic giant branch star of
radius $\sim$600 $R_{\odot}$, far exceeding its current binary separation
$\sim$$3\,R_{\odot}$.  He proposed that the dissipation of orbital energy
provided the means both for planetary nebula ejection and for the severe
orbital contraction between initial and final states, a process he labeled
`common envelope evolution' (not to be confused with the common envelopes of
contact binary stars).  Discovery soon followed of the first `smoking gun',
the short-period eclipsing nucleus of the planetary nebula
Abell~63~\cite{Bon78}.

Over the succeeding three decades, there have been a number of attempts to
build detailed physical models of common envelope evolution (see~\cite{Taa00}
for a review).  These efforts have grown significantly in sophistication, but
this phenomenon presents a daunting numerical challenge, as common envelope
evolution is inherently three-dimensional, and the range of spatial and
temporal scales needed to represent a common envelope binary late in its
inspiral can both easily exceed factors of $10^3$.  Determining the efficiency
with which orbital energy is utilized in envelope ejection requires such a
code to conserve energy over a similarly large number of dynamical time
scales.

Theoretical models of common envelope evolution are not yet capable of
predicting the observable properties of objects in the process of inspiral. 
If envelope ejection is to be efficient, then the bulk of dissipated orbital
energy must be deposited in the common envelope on a time scale short compared
with the thermal time scale of the envelope, else that energy be lost to
radiation.  The duration of the common envelope phase thus probably does not
exceed $\sim\!\!10^3$ years.  However, general considerations of the high
initial orbital angular momenta of systems such as the progenitor of V471 Tau,
and the fact that most of the orbital energy is released the envelope only
very late in the inspiral have led to a consensus
view~\cite{Web79,Mor81,Mor87,Yor95,Sok98} that the planetary nebulae they
eject should be bipolar in structure, with dense equatorial rings absorbing
most of the initial angular momentum of the binary, and higher-velocity polar
jets powered by the late release of orbital energy.  Indeed, this appears to
be a signature morphology of planetary nebulae with binary nuclei
(e.g.,~\cite{Bon00}), although it may not be unique to binary nuclei.

\section{The Energetics of Common Envelope Evolution}
\label{Energetics}

Notwithstanding the difficulties in modeling common envelope evolution in
detail, it is possible to calculate with some confidence the initial total
energy and angular momentum of a binary at the onset of mass transfer, and the
corresponding orbital energy and angular momentum of any putative remnant of
common envelope evolution.  

Consider an initial binary of component masses $M_1$ and $M_2$, with orbital
semimajor axis $A_{\rm i}$.  Its initial total orbital energy is
\begin{equation}
\label{Eorbi}
E_{\rm orb,i} = - \frac{GM_1 M_2}{2 A_{\rm i}} \ .
\end{equation}
Let star 1 be the star that initiates interaction upon filling its Roche lobe. 
If $M_{\rm 1c}$ is its core mass, and $M_{\rm 1e} = M_1 - M_{\rm 1c}$ its
envelope mass, then we can write the initial total energy of that envelope as
\begin{equation}
\label{Eenv}
E_{\rm e} = - \frac{GM_1 M_{\rm 1e}}{\lambda R_{\rm 1,L}} \ ,
\end{equation}
where $R_{\rm 1,L}$ is the Roche lobe radius of star 1 at the onset of mass
transfer (the orbit presumed circularized prior to this phase), and $\lambda$
is a dimensionless parameter dependent on the detailed structure of the
envelope, but presumably of order unity.  For very simplified models of red
giants -- condensed polytropes~\cite{Ost53,Hae55,Hje87} -- $\lambda$ is a
function only of $m_{\rm e} \equiv M_{\rm e}/M = 1 - M_{\rm c}/M$, the ratio
of envelope mass to total mass for the donor, and is well-approximated by
\begin{equation}
\label{lambda}
\lambda^{-1} \approx 3.000 - 3.816 m_{\rm e} + 1.041 m_{\rm e}^2 + 0.067
m_{\rm e}^3 + 0.136 m_{\rm e}^4 \ ,
\end{equation}
to within a relative error $<10^{-3}$.

For the final orbital energy of the binary we have
\begin{equation}
\label{Eorbf}
E_{\rm orb,f} = - \frac{GM_{\rm 1c} M_2}{2 A_{\rm f}} \ ,
\end{equation}
where $A_{\rm f}$ is of course the final orbital separation.  If a fraction
$\alpha_{\rm CE}$ of the difference in orbital energy is consumed in unbinding
the common envelope,
\begin{equation}
\label{alphaCE}
\alpha_{\rm CE} \equiv \frac{E_{\rm e}}{(E_{\rm orb}^{\rm (f)} - E_{\rm
orb}^{\rm (i)})} \ ,
\end{equation}
then
\begin{equation}
\label{Afalpha}
\frac{A_{\rm f}}{A_{\rm i}} = \frac{M_{\rm 1c}}{M_1} \left[ 1 + \left(
\frac{2}{\alpha_{\rm CE} \lambda r_{\rm 1,L}} \right) \left( \frac{M_1 -
M_{\rm 1c}}{M_2} \right) \right]^{-1} \ ,
\end{equation}
where $r_{\rm 1,L} \equiv R_{\rm 1,L}/A_{\rm i}$ is the dimensionless Roche
lobe radius of the donor at the start of mass transfer.  In the classical
Roche approximation, $r_{\rm 1,L}$ is a function only of the mass ratio, $q
\equiv M_1/M_2$~\cite{Egg83}:
\begin{equation}
\label{Roche}
r_{\rm 1,L} \approx \frac{0.49 q^{2/3}}{0.6 q^{2/3} + \ln (1 + q^{1/3})} \ .
\end{equation}
Typically, the second term in brackets in (\ref{Afalpha}) dominates the first
term.

As formulated above, our treatment of the outcome of common envelope evolution
neglects any sources or sinks of energy beyond gravitational terms and the
thermal energy content of the initial envelope (incorporated in the parameter
$\lambda$).  The justification for this assumption is again that common
envelope evolution must be rapid compared to the thermal time scale of the
envelope.  This implies that radiative losses (or nuclear energy gains -- see
below) are small.  They, as well as terminal kinetic energy of the ejecta, are
presumably reflected in ejection efficiencies $\alpha_{\rm CE} < 1$.  We
neglect also the rotational energy of the common envelope (invariably small in
magnitude compared to its gravitational binding energy), and treat the core of
the donor star and the companion star as inert masses, which neither gain nor
lose mass or energy during the course of common envelope evolution.  One might
imagine it possible that net accretion of mass by the companion during
inspiral might compromise this picture.  However, the common envelope is
typically vastly less dense than the companion star. and may be heated to
roughly virial temperature on infall.  A huge entropy barrier arises at the
interface between the initial photosphere of the companion and the common
envelope in which it is now embedded, with a difference in entropy per
particle of order $(\mu m_{\rm H}/k) \Delta s \approx$ 4--6.  The rapid rise
in temperature and decrease in density through the interface effectively
insulates the accreting companion thermally, and strongly limits the fraction
of the very rarified common envelope it can retain upon exit from that
phase~\cite{Web88,Hje91}.

Common envelope evolution entails systemic angular momentum losses as well as
systemic mass and energy losses.  Writing the orbital angular momentum of the
binary,
\begin{equation}
\label{Jorb}
J = \left[ \frac{GM_1^2 M_2^2 A (1 - e^2)}{M_1 + M_2} \right]^{1/2} \ ,
\end{equation}
in terms of the total orbital energy, $E = -GM_1 M_2/2A$, we find immediately
that the ratio of final to initial orbital angular momentum is
\begin{equation}
\label{JfJi}
\frac{J_{\rm f}}{J_{\rm i}} = \left( \frac{M_{\rm 1c}}{M_1} \right)^{3/2}
\left( \frac{M_{\rm 1c} + M_2}{M_1 + M_2} \right)^{-1/2} \left( \frac{E_{\rm
i}}{E_{\rm f}} \right)^{1/2} \left( \frac{1 - e_{\rm f}^2}{1 - e_{\rm i}^2}
\right)^{1/2} \ .
\end{equation}
Since $M_{\rm 1c} < M_1$ and we expect the initial orbital eccentricity to be
small ($e_{\rm i} \approx 0$), it follows that any final energy state lower
than the initial state ($|E_{\rm f}| > |E_{\rm i}|$) requires the loss of
angular momentum.  The reverse is not necessarily true, so it is the energy
budget that most strongly constrains possible outcomes of common envelope
evolution.

\section{Does Common Envelope Evolution Work?}
\label{Work}

As an example of common envelope energetics, let us revisit the pre-CV
V471~Tau, applying the simple treatment outlined above.  It is a member of the
Hyades, an intermediate-age metal-rich open cluster ($t = 650$ Myr, [Fe/H] =
+0.14) with turnoff mass $M_{\rm TO} = 2.60 \pm 0.06\,M_{\odot}$~\cite{Leb01}. 
The cooling age of the white dwarf is much smaller than the age of the cluster
($t_{\rm cool,WD} = 10^7$ yr ~\cite{OBr01} -- but see the discussion there of
the paradoxical fact that this most massive of Hyades white dwarfs is also the
youngest).  Allowing for the possibility of significant mass loss in a stellar
wind prior to the common envelope phase, we may take $M_{\rm TO}$ for an upper
limit to the initial mass $M_1$ of the white dwarf component.  The current
masses for the white dwarf and its dK2 companion, as determined by O'Brien et
al.~\cite{OBr01} are $M_{\rm WD} = 0.84 \pm 0.05\,M_{\odot}$, $M_{\rm K} =
0.93 \pm 0.07\,M_{\odot}$, with orbital separation $A = 3.30 \pm
0.08\,R_{\odot}$.  A $2.60\,M_{\odot}$ star of Hyades metallicity with a
$0.84\,M_{\odot}$ core lies on the thermally-pulsing asymptotic giant branch,
with radius (maximum in the thermal pulse cycle) which we estimate at $R_{\rm
i} = 680\,R_{\odot} = R_{\rm 1,L}$, making $A_{\rm i} = 1450\,R_{\odot}$. 
With this combination of physical parameters, we derive an estimate of
$\alpha_{\rm CE} \lambda = 0.057$ for V471 Tau.  Equation~(\ref{lambda}) then
implies $\alpha_{\rm CE} = 0.054$.  This estimate of course ignores any mass
loss prior to common envelope evolution (which would drive $\alpha_{\rm CE}$
to lower values), or orbital evolution since common envelope evolution (which
would drive $\alpha_{\rm CE}$ to higher values).  In any event, the status of
V471~Tau would appear to demand only a very small efficiency of envelope
ejection.\footnote{The anomalously small value of $\alpha_{\rm CE}$ deduced
for V471 Tau may be connected to its puzzlingly high white dwarf mass and
luminosity:  O'Brien et al.~\cite{OBr01} suggest that it began as a
heirarchical triple star, in which a short-period inner binary evolved into
contact, merged (as a blue straggler), and later engulfed its lower-mass
companion in a common envelope.  An overmassive donor at the onset of common
envelope evolution would then have a more massive core than produced by its
contemporaries among primordially single stars, and it would fill its Roche
lobe with a more massive envelope at somewhat shorter orbital period, factors
all consistent with a larger value of $\alpha_{\rm CE}$ having led to V471 Tau
as now observed.}

The fact that V471 Tau is a double-lined eclipsing member of a well-studied
cluster provides an exceptionally complete set of constraints on its prior
evolution.  In all other cases of short-period binaries with degenerate or
compact components, available data are inadequate to fix simultaneously both
the initial mass of the compact component and the initial binary separation,
for example.  To validate the energetic arguments outlined above, one must
resort to consistency tests, whether demonstrating the existence of
physically-plausible initial conditions that could produce some individual
system, or else following a plausible distribution of primordial binaries
wholesale through the energetics of common envelope evolution and showing
that, after application of appropriate observational selection effects, the
post-common-envelope population is statistically consistent with the observed
statistics of the selected binary type.  In the cases of interacting binaries,
such as CVs, one should allow further for post-common-envelope evolution. 
Nevertheless, within these limitations, binary population synthesis models
show broad consistency between the outcomes of common envelope evolution and
the statistical properties of CVs and
pre-CVs~\cite{deK92,Kol93,Pol96,How01,Wil04}, as well as with most super-soft
X-ray sources~\cite{DiS94}, for assumed common envelope ejection efficiencies
typically of order $\alpha_{\rm CE} \approx 0.3$--0.5.

\begin{FPfigure}
\setlength\fboxsep{0pt}
\centering
\resizebox{.97\textwidth}{!}{\includegraphics{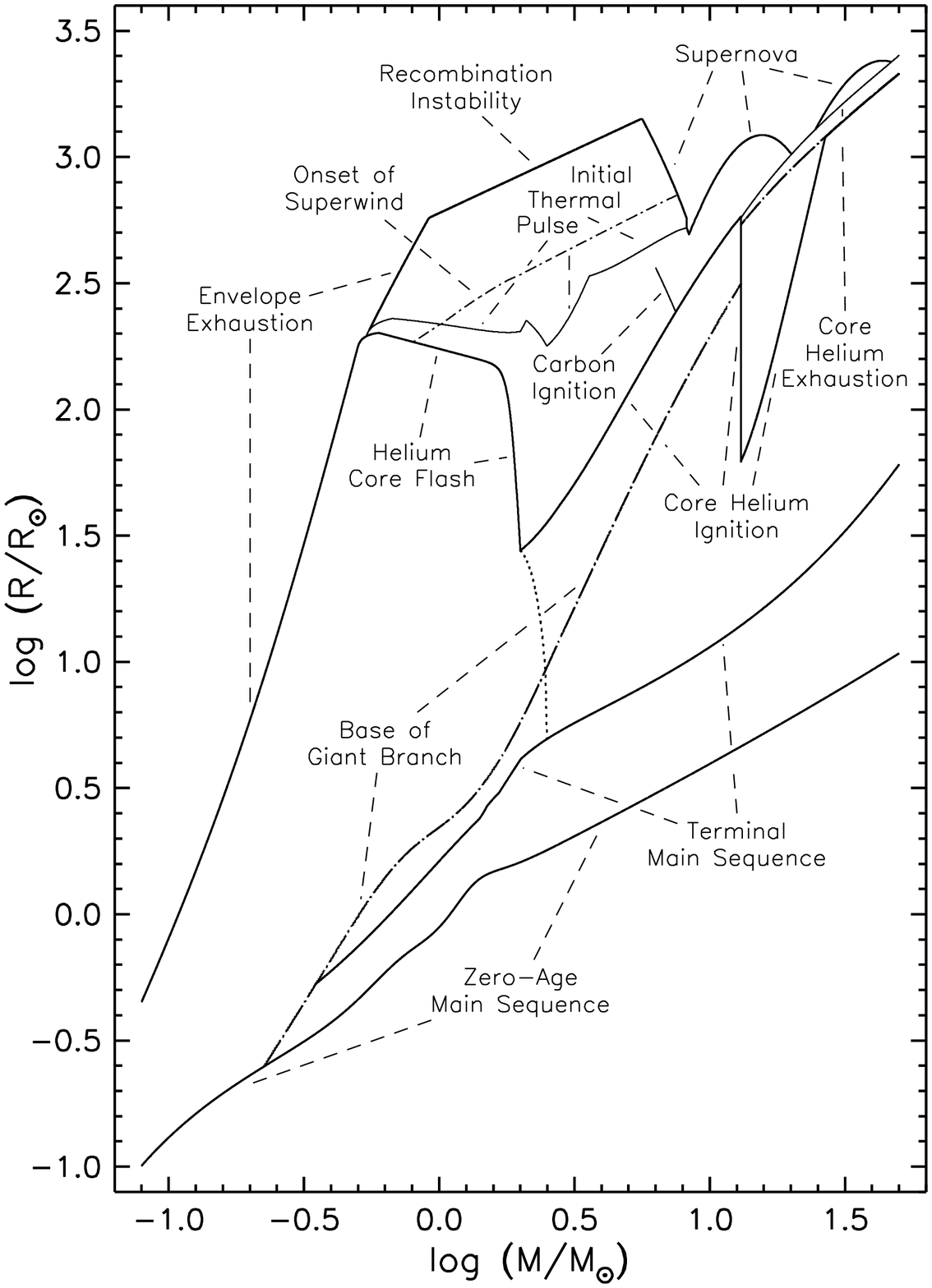}}
\caption{The mass-radius diagram for stars of solar metallicity, constructed
from the parametric models of stellar evolution by Hurley, Pols, \&
Tout~\cite{Hur00} and models of thermally-pulsing asymptotic giant branch
stars by Wagenhuber \& Weiss~\cite{Wag94}.  Also plotted in the locus of
asymptotic giant branch stars at the onset of the superwind, after
Willson~\cite{Wil00}; beyond this radius, systemic mass loss drives orbital
expansion faster than nuclear evolution drives stellar expansion, and a binary
will no longer be able to initiate tidal mass transfer.  The unlabeled dotted
line terminating at the junction between lines labeled `helium core flash' and
`core helium ignition' marks the division between those helium cores (at lower
masses) which evolve to degeneracy if stripped of their envelope, and those
(at higher masses) which ignite helium non-degenerately and become helium
stars.}
\label{MRdiagram}
\end{FPfigure}

A useful tool in reconstructing the evolutionary history of a binary, used
implicitly above in analyzing V471~Tau, is the mass-radius diagram spanned by
single stars of the same composition as the binary.  Figure~\ref{MRdiagram}
illustrates such a diagram for solar-composition stars from $0.08\ M_{\odot}$
to $50\ M_{\odot}$.  In it are plotted various critical radii marking, as a
function of mass, the transition from one evolutionary phase to the
next.\footnote{Not all evolutionary phases are represented here.  In a binary,
a donor initiates mass transfer when it first fills its Roche lobe; if it
would have done so at a prior stage of evolution, then its present
evolutionary state is `shadowed', in the sense that it only occurs by virtue
of the binary \emph{not} having filled its lobe previously.  Thus, for
example, low- and intermediate-mass stars cannot in general initiate mass
transfer during core helium burning, because they would have filled their
Roche lobes on the initial ascent of the giant branch.}  Since the Roche lobe
of a binary component represents a dynamical limit to its size, its orbital
period fixes the mean density at which that star fills its Roche lobe,
\begin{equation}
\label{Porb}
\log P_{\rm orb}({\rm d}) \approx \frac{3}{2} \log (R_{\rm L}/R_{\odot}) -
\frac{1}{2} \log (M/M_{\odot}) - 0.455 \ ,
\end{equation}
to within a very weak function of the binary mass ratio.  The mass and radius
of any point in Fig.~\ref{MRdiagram} therefore fixes the orbital period at
which such a star would fill its Roche lobe, just as the orbital period of a
binary fixes the evolutionary state at which such a star initiates mass
transfer.

\begin{FPfigure}
\setlength\fboxsep{0pt}
\centering
\resizebox{.97\textwidth}{!}{\includegraphics{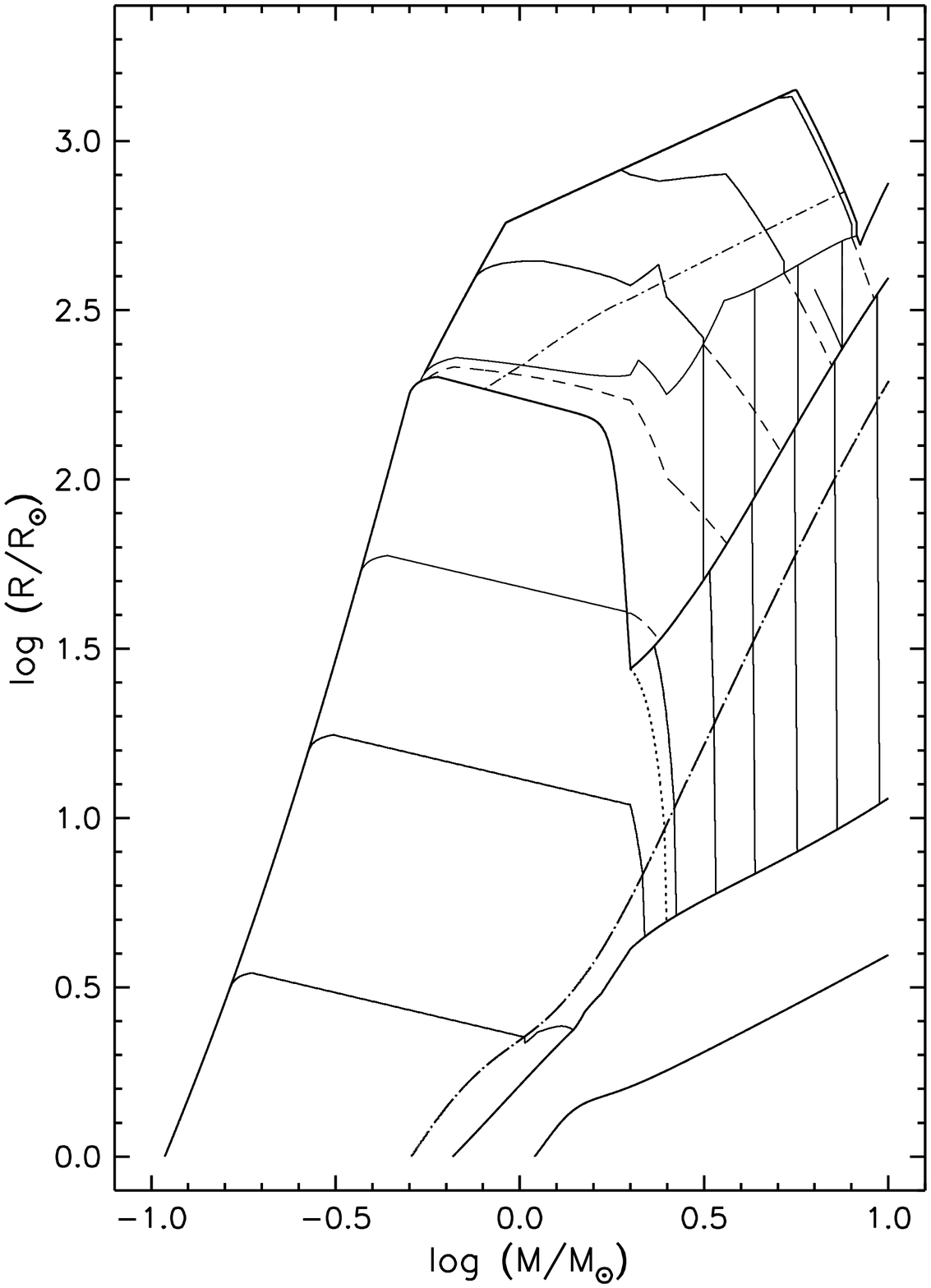}}
\caption{The mass-radius diagram for low- and intermediate-mass stars, as in
Fig.~\ref{MRdiagram}, but with loci of constant core mass added.  The solid
lines added correspond to core masses interior to the hydrogen-burning shell,
dashed lines to those interior to the helium-burning shell.  Solid lines
intersecting the base of the giant branch (dash-dotted curve) correspond to
helium core masses of to 0.15, 0.25, 0.35, 0.5, 0.7, 1.0, 1.4, and 2.0
$M_{\odot}$; those between helium ignition and the initial thermal pulse to
0.7, 1.0, 1.4, and 2.0 $M_{\odot}$, and those beyond the initial thermal pulse
to 0.7, 1.0, and 1.4 $M_{\odot}$.  Dashed lines between helium ignition and
initial thermal pulse correspond to carbon-oxygen core masses of 0.35, 0.5,
0.7, 1.0, and 1.4 $M_{\odot}$.  Beyond the initial thermal pulse, helium and
carbon-oxygen core masses converge, with the second dredge-up phase reducing
helium core masses above $\sim$0.8 $M_{\odot}$ to the carbon-oxygen core.}
\label{Mcore_diagram}
\end{FPfigure}

In Fig.~\ref{Mcore_diagram}, the corresponding core masses of low- and
intermediate-mass stars are plotted in the mass-radius diagram.  For a binary
which is the immediate product of common envelope evolution, the mass of the
most recently formed white dwarf (presumably the spectroscopic primary) equals
the core mass of the progenitor donor star.  That donor (presuming it to be of
solar metallicity) must be located somewhere along the corresponding core mass
sequence in Fig.~\ref{Mcore_diagram}, with the radius at any point along that
sequence corresponding to the Roche lobe radius at the onset of the mass
transfer, and the mass a that point corresponding to the initial total mass of
the donor.  Thus, if the mass of the most recently-formed white dwarf is
known, it is possible to identify a single-parameter (e.g., initial mass or
initial radius of the donor) family of possible common-envelope progenitors.

Using a mapping procedure similar to this, Nelemans \& Tout~\cite{Nel05}
recently explored possible progenitors for detached close binaries with white
dwarf components.  Broadly speaking, they found solutions
using~(\ref{Afalpha}) for almost all systems containing only one white dwarf
component.  Only three putative post-common-envelope systems failed to yield
physically-plausible values of $\alpha_{\rm CE} \lambda$: AY~Cet (G5~III + DA,
$P_{\rm orb} = 56.80\,{\rm  d}$~\cite{Sim85}), Sanders~1040 (in M67: G4~III +
DA, $P_{\rm orb} = 42.83\,{\rm d}$~\cite{vdB99}), and HD~185510 (=V1379~Aql:
gK0 + sdB, $P_{\rm orb} = 20.66\,{\rm d}$~\cite{Jef97}).  The first two of
these systems are non-eclipsing, but photometric masses for their white dwarf
components are extremely low (estimated at $\sim\!\!0.25\,M_{\odot}$ and
$0.22\,M_{\odot}$, respectively), with Roche lobe radii consistent with the
limiting radii of very low-mass giants as they leave the giant branch (cf.
Fig.~\ref{Mcore_diagram}, above).  They are thus almost certainly post-Algol
binaries, and not post-common-envelope binaries.  HD~185510 is an eclipsing
binary; a spectroscopic orbit exists only for the gK0 component~\cite{Fek93}. 
The mass ($0.304 \pm 0.015\,M_{\odot}$) and radius ($0.052 \pm
0.010\,R_{\odot}$) of the sdB component, deduced from model atmosphere fitting
of IUE spectra combined with solution of the eclipse light curve, place it on
a low-mass white dwarf cooling curve, rather than among helium-burning
subdwarfs~\cite{Jef97}.  Indeed, from fitting very detailed evolutionary
models to this system, Nelson \& Eggleton~\cite{NeE01} found a post-Algol
solution they deemed acceptable.  It thus appears that these three problematic
binaries are products of quasi-conservative mass transfer, and not common
envelope evolution.

The close double white dwarfs present a more difficult conundrum, however. 
Nelemans et al.~\cite{Nel00,Nel01,Nel05} found it impossible using the
energetic arguments~(\ref{Afalpha}) outlined above to account for the
existence of a most known close double white dwarfs.  Mass estimates can be
derived for spectroscopically detectable components of these systems from
their surface gravities and effective temperatures (determined from Balmer
line fitting).  The deduced masses are weakly dependent on the white dwarf
composition, and may be of relatively modest accuracy, but they are
independent of the uncertainties in orbital inclination afflicting orbital
solutions.  These mass estimates place the great majority of detectable
components in close double white dwarf binaries below
$\sim\!\!0.46\,M_{\odot}$, the upper mass limit for pure helium white dwarfs
(e.g.,~\cite{Swe90}).  They are therefore pure helium white dwarfs, or perhaps
hybrid white dwarfs (low-mass carbon-oxygen cores with thick helium
envelopes).  While reconstructions of their evolutionary history yield
physically-reasonable solutions for the final common envelope phase, with
values for $0 < \alpha_{\rm CE} \lambda < 1 $, the preceding phase of mass
transfer, which gave rise to the first white dwarf, is more problematic.  If
it also proceeded through common envelope evolution, the deduced values of
$\alpha_{\rm CE} \lambda \le -4$ for that phase are unphysical.  Nelemans \&
Tout~\cite{Nel05} interpreted this paradox as evidence that descriptions of
common envelope evolution in terms of orbital energetics, as described above,
are fundamentally flawed.

\section{An Alternative Approach to Common Envelope Evolution?}
\label{Alternative}

Nelemans et al.~\cite{Nel00} proposed instead parameterizing common envelope
evolution in terms of $\gamma$, the ratio of the fraction of angular momentum
lost to the fraction of mass lost:
\begin{equation}
\label{gamma}
\frac{J_{\rm i} - J_{\rm f}}{J_{\rm i}} = \gamma \frac{M_1 - M_{\rm 1,c}}{M_1
+ M_2}\ .
\end{equation}
Both initial and final orbits are assumed circular, so the ratio of final to
initial orbital separations becomes
\begin{equation}
\label{Afgamma}
\frac{A_{\rm f}}{A_{\rm i}} = \left( \frac{M_1}{M_{\rm 1c}} \right)^2 \left(
\frac{M_{\rm 1c} + M_2}{M_1 + M_2} \right) \left[ 1 - \gamma \left( \frac{M_1
- M_{\rm 1c}}{M_1 + M_2} \right) \right]^2\ .
\end{equation}
Among possible solutions leading to known close double white dwarfs, Nelemans
\& Tout~\cite{Nel05} find values $1 < \gamma \lesssim 4$ required for the
second (final) common envelope phase, and $0.5 \lesssim \gamma < 3$ for the
first (putative) common envelope phase.  They note that values in the range
$1.5 < \gamma < 1.7$ can be found among possible solutions for all common
envelope phases in their sample, not only those leading to known double white
dwarfs, but those leading to known pre-CV and sdB binaries as well.

The significance of this finding is itself open to debate.  At one extreme, it
would seem implausible for any mechanism to remove less angular momentum per
unit mass than the orbital angular momentum per unit mass of either component
in its orbit (so-called Jeans-mode mass loss).  At the other extreme, a firm
upper limit to $\gamma$ is set by vanishing final orbital angular momentum,
$J_{\rm f}$.  If $M_{\rm 1c}$ and $M_2$ can be regarded as fixed, the
corresponding limits on $\gamma$ are
\begin{equation}
\label{gammalimits}
\left( \frac{M_1 + M_2}{M_1 - M_{\rm 1c}} \right) > \gamma > \left( \frac{M_1
+ M_2}{M_1 - M_{\rm 1c}} \right) \left[ 1 - \left( \frac{M_{\rm 1c}}{M_1}
\right) \left( \frac{M_1 + M_2}{M_{\rm 1c} + M_2} \right) \right] \ .
\end{equation}
In a fairly typical example, $M_{\rm 1c} = M_2 = \frac{1}{4} M_1$, $\gamma$ is
inevitably tightly constrainted for any conceivable outcome: $\frac{5}{3} >
\gamma > \frac{5}{8}$.  The ratio of final to initial orbital separation,
$A_{\rm f}/A_{\rm i}$, is extremely sensitive to $\gamma$ near the upper limit
of its range.  It is therefore not surprising to find empirical estimates of
$\gamma$ clustering as they do -- their values merely affirm the fact that
$A_{\rm f}$ must typically be much smaller than $A_{\rm i}$.

The unphysically large or, more commonly, \emph{negative} values of
$\alpha_{\rm CE} \lambda$ noted above for the first mass transfer phase in the
production of close white dwarf binaries~\cite{Nel05} implies that the orbital
energies of these binaries have \emph{increased} through this phase (or, at
any rate, decreased by significantly less than the nominal binding energies of
their common envelopes).  Such an increase in orbital energy is a hallmark of
slow, quasi-conservative mass transfer, on a thermal or nuclear time scale. 
Thermal time scale mass transfer is driven by relaxation of the donor star
toward thermal evolution; the re-expansion of the donor following mass ratio
reversal is powered by the (nuclear) energy outflow from the core of the star. 
Likewise, the bulk expansion of the donor star in nuclear time scale mass
transfer draws energy from nuclear sources in that star.  It appears,
therefore, that the first phase of mass transfer among known close double
white dwarfs cannot have been a common envelope phase, but must instead have
been a quasi-conservative phase, notwithstanding the difficulties that
conclusion presents, as we shall now see.

\begin{FPfigure}
\setlength\fboxsep{0pt}
\centering
\resizebox{.97\textwidth}{!}{\includegraphics{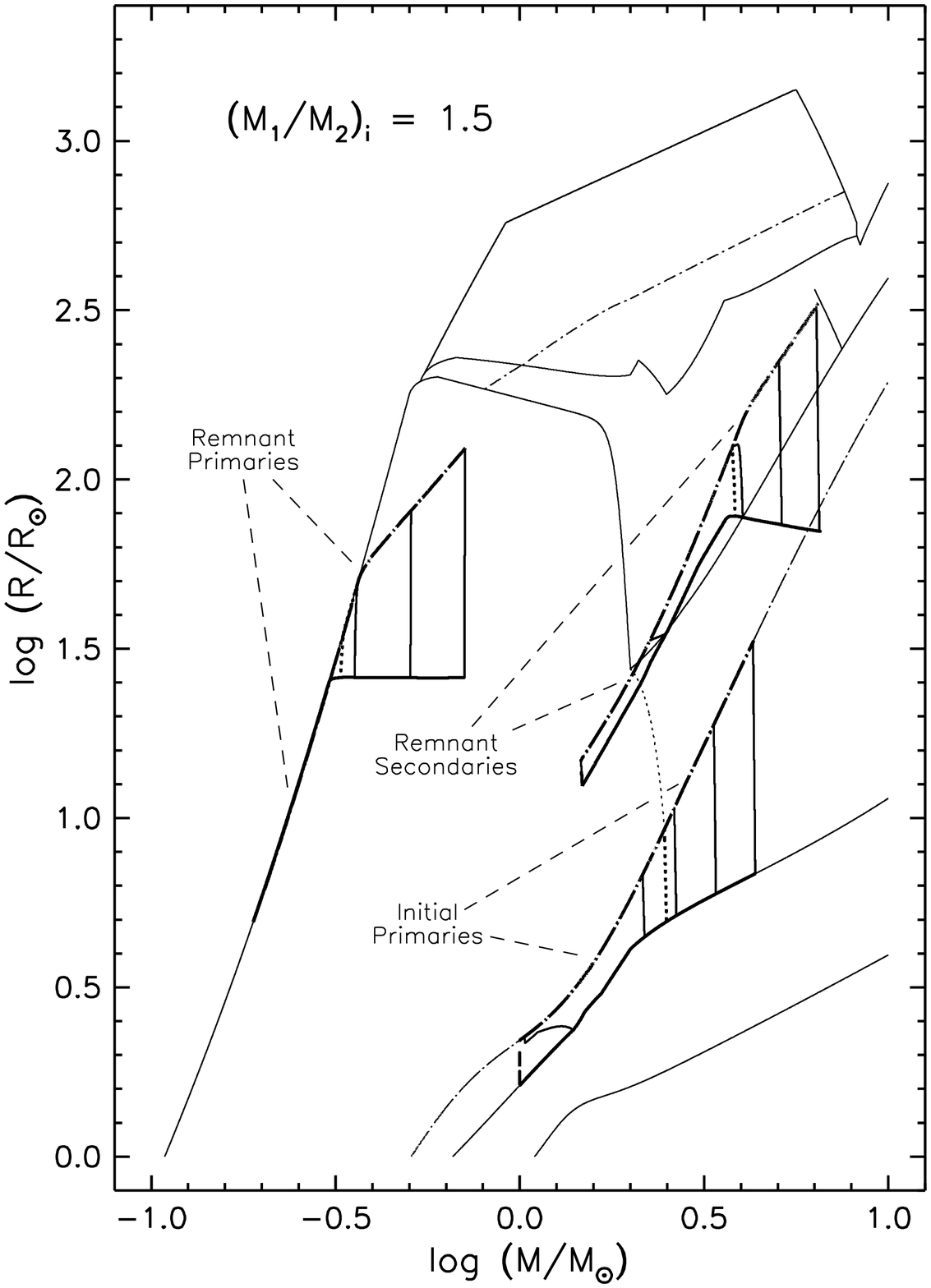}}
\caption{Products of mass- and angular momentum-conservative mass transfer for
a typical initial mass ratio.  The radii indicated refer to \emph{Roche lobe
radii} at the onset or termination of mass transfer, as appropriate.  To avoid
common envelope evolution, the donor stars (the region outlined in bold toward
the lower right in the diagram) must have radiative envelopes, and so arise
between the terminal main sequence and base of the giant branch.  Their mass
transfer remnants are outlined in bold at the center-left of the diagram, with
the remnant accretors at upper right.  The regions mapped are truncated in
each case at a lower initial donor mass of 1.0~$M_{\odot}$ and upper initial
donor core mass of 0.7~$M_{\odot}$.  Lines of constant initial core mass (with
values as in Fig.~\ref{Mcore_diagram}) are indicated for the initial and
remnant primaries.  Lines of constant remnant primary mass are indicated for
the remnant secondaries.}
\label{ConsvMT}
\end{FPfigure}

The dilemma that the close double white dwarfs present is illustrated in
Figs.~\ref{ConsvMT} and ~\ref{Phase2}.  Figure~\ref{ConsvMT} shows the
distribution of immediate remnants of mass transfer among solar-metallicity
binaries of low and intermediate mass, for a relatively moderate initial mass
ratio.  Conservation of total mass and orbital angular momentum have been
assumed.  The remnants of the intial primary include both degenerate helium
white dwarfs, and nondegenerate helium stars which have lost nearly all of
their hydrogen envelopes.  The helium white dwarfs lie almost entirely along
the left-hand boundary, the line labeled `envelope exhaustion' in
Fig.~\ref{MRdiagram}.  (The extent of this sequence is more apparent in the
distribution of remnant secondaries.)  Their progenitors have enough angular
momentum to accommodate core growth in the terminal phases of mass transfer. 
In the calculation shown, the least massive cores grow from $0.11\,M_{\odot}$
to $\sim\!\!0.18\,M_{\odot}$ by the completion of mass transfer.  In contrast,
virtually all binaries leaving nondegenerate helium star remnants have too
little angular momentum to recover thermal equilibrium before they have lost
their hydrogen envelopes; for them, there is no slow nuclear time scale phase
of mass transfer, and core growth during mass transfer is negligible.  The
lowest-mass helium star remnants have nuclear burning lifetimes comparable to
their hydrogen-rich binary companions, now grown through mass accretion. 
Those more massive than $\sim\!\!0.8\,M_{\odot}$ develop very extended
envelopes during shell helium burning, and will undergo a second phase of mass
transfer from primary to secondary, not reflected here; such massive white
dwarfs are absent in the Nelemans \& Tout~\cite{Nel05} sample, and so are
omitted here.

\begin{FPfigure}
\setlength\fboxsep{0pt}
\centering
\resizebox{.97\textwidth}{!}{\includegraphics{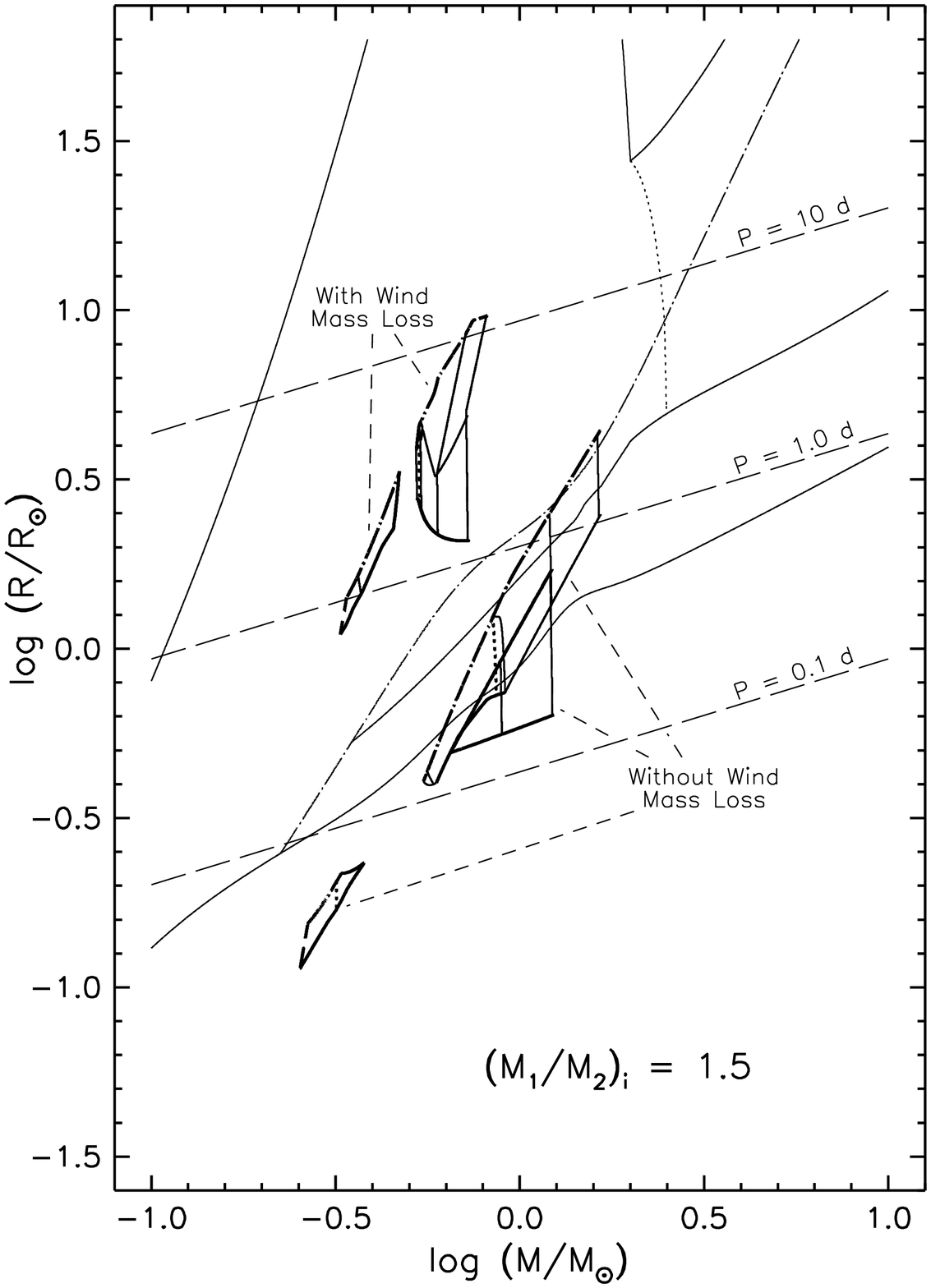}}
\caption{Remnants of the second, common envelope, phase of mass transfer of
the systems shown in Fig.~\ref{ConsvMT}.  Masses refer to the final remnants
of the original secondaries, and radii to their Roche lobe radii.  Two groups
of remnants are shown.  Those at lower right labeled `Without Wind Mass Loss'
follow directly from the distributions of remnant primaries and secondaries
shown in Fig.~\ref{ConsvMT}.  Because the remnant secondaries straddle the
helium ignition line in Fig.~\ref{ConsvMT}, across which core masses are
discontinuous (see Fig.~\ref{Mcore_diagram}), the distribution of their
post-common-envelopes remnants is fragmented, some appearing as degenerate
helium white dwarf remnants (lower left), some as helium main sequence star
remnants (lower center), and the remainder as shell-burning helium star
remnants (upper center).  These latter two groups overlap in the mass-radius
diagram.  The remnant distributions labeled `With Wind Mass Loss' assume that
the remnant secondaries of conservative mass transfer lose half their mass in
a stellar wind prior to common envelope evolution.  They too are fragmented,
into degenerate helium white dwarf remnants (lower left) and shell-burning
helium stars (upper right).  Within each group of remnants, lines of constant
remnant primary mass are shown, as in Fig.~\ref{ConsvMT}.}
\label{Phase2}
\end{FPfigure}

In Fig.~\ref{Phase2}, the remnants of the first phase of conservative mass
transfer illustrated in Fig.~\ref{ConsvMT} are followed through the second
phase of mass transfer, using~(\ref{Afalpha}).  Because the remnants of the
first phase have second-phase donors much more massive their companions, and
nearly all have deep convective envelopes, they are unstable to dynamical time
scale mass transfer, and undergo common envelope evolution.  The systems
labeled `Without Wind Mass Loss' have been calculated assuming that no orbital
evolution or mass loss occurs between the end of the first phase of mass
transfer and common envelope evolution.  It is assumed furthermore that
$\alpha_{\rm CE} = 1$, in principle marking the most efficient envelope
ejection energetically possible.  Binary orbital periods of 0.1, 1.0, and 10
days (assuming equal component masses) are indicated for reference.

Observed close double white dwarfs, as summarized by Nelemans \&
Tout~\cite{Nel05}, have a median orbital period of $1.4\,{\rm d}$, and mass
(spectroscopic primary) $0.39\,M_{\odot}$.  Among double-lines systems, nearly
equal white dwarf masses are strongly favored, with the median $q = 1.00$. 
Clearly, most observed double white dwarfs are too long in orbital period
(have too much total energy and angular momentum) to have evolved in the
manner assumed here.  Furthermore, the computed binary mass ratios are
typically more extreme than observed, with the second-formed core typically
1.3–-2.5 times as massive as the first.  The problem is that, while remnant
white dwarfs or low-mass helium stars with suitable masses can be produced in
the first, conservative mass transfer phase, the remnant companions have
envelope masses too large, and too tightly bound, to survive the second
(common envelope) phase of interaction at orbital separations and periods as
large as observed.  Evidently, the progenitors of these double white dwarfs
have lost a significant fraction of their initial mass, while gaining in
orbital energy, prior to the final common envelope phase.  These requirements
can be fulfilled by a stellar wind, provided that the process is slow enough
that energy losses in the wind can be continuously replenished from nuclear
energy sources.

The requisite mass loss and energy gain are possible with stellar wind mass
loss during the non-interactive phase between conservative and common envelope
evolution, or with stellar winds in nuclear time-scale mass transfer or the
terminal (recovery) phase of thermal time-scale mass transfer.  Systemic mass
loss during or following conservative mass transfer will (in the absence of
angular momentum losses) shift the remnant regions to the left and upward in
Fig.~\ref{ConsvMT} (subject to the limit posed by envelope exhaustion), while
systemic angular momentum losses shift them downwards.  More extreme initial
mass ratios shift them downwards to the left.

The net effect of wind mass loss is illustrated by the regions labeled `With
Wind Mass Loss' in Fig.~\ref{Phase2}.  For simplicity, it is assumed here that
half the remnant mass of the original secondary was lost in a stellar wind
prior to the common envelope phase.  Mass loss on this scale not only
significantly reduces the mass of the second-formed white dwarf relative to
the first, but the concomitant orbital expansion produces wider remnant double
white dwarfs, bringing this snapshot model into good accord with the general
properties of real systems.  Losses of this magnitude might be unprecedented
among single stars prior to their terminal superwind phase, but they have been
a persistent feature of evolutionary studies of Algol-type binaries~\cite[and
references therein]{Giu81} and, indeed, of earlier studies of close double
white dwarf formation~\cite{Han98}.  In the present context, their existence
appears inescapable, if not understood.

\section{Long-Period Post-Common-Envelope Binaries and the Missing Energy
Problem}
\label{MissingEnergy}

If the properties of short-period binaries with compact components can be
reconciled with the outcomes of common envelope evolution as expected from
simple energetics arguments, a challenge to this picture still comes from the
survival of symbiotic stars and recurrent novae at orbital separations too
large to have escaped tidal mass transfer earlier in their evolution. 
Notwithstanding this author's earlier hypothesis that the outbursting
component in the recurrent nova T~CrB (and its sister system RS~Oph) might be
a nondegenerate star undergoing rapid accretion~\cite{Web76,Liv86,Web87}, it
is now clear that the hot components in both of these systems must indeed be
hot, degenerate dwarfs~\cite{Sel92,Dob96,Bel98,Hri98}.  Furthermore, the short
outburst recurrence times of these two binaries demand that the degenerate
dwarfs in each must have masses very close to the Chandrasekhar limit.

The complexion of the problem posed by these systems can be illustrated by a
closer examination of T~CrB itself.  Its orbital period (P = 227.53 d) and
spectroscopic mass function ($f(m) = 0.299 M_{\odot}$) are well-established
from the orbit of the donor M3~III star~\cite{Ken86}.  The emission-line orbit
for the white dwarf~\cite{Kra58} now appears very doubtful~\cite{Hri98}, but
the system shows very strong ellipsoidal variation (e.g.,~\cite{Bel98}),
suggesting that the system is near a grazing eclipse.  Following Hric, et
al.~\cite{Hri98}, I adopt $M_{\rm WD} = 1.38\ M_{\odot}$ amd $M_{\rm M3} =
1.2\ M_{\odot}$.  The Roche lobe radius of the white dwarf is then $R_{\rm
L,WD} = 84\ R_{\odot}$, nearly an order of magnitude larger than can be
accommodated from the energetics arguments presented above, even for
$\alpha_{\rm CE} = 1$, assuming solar metallicity for the system (see
Fig.~\ref{PostCEalpha}).  A similar discrepancy occurs for RS~Oph.

\begin{FPfigure}
\setlength\fboxsep{0pt}
\centering
\resizebox{.97\textwidth}{!}{\includegraphics{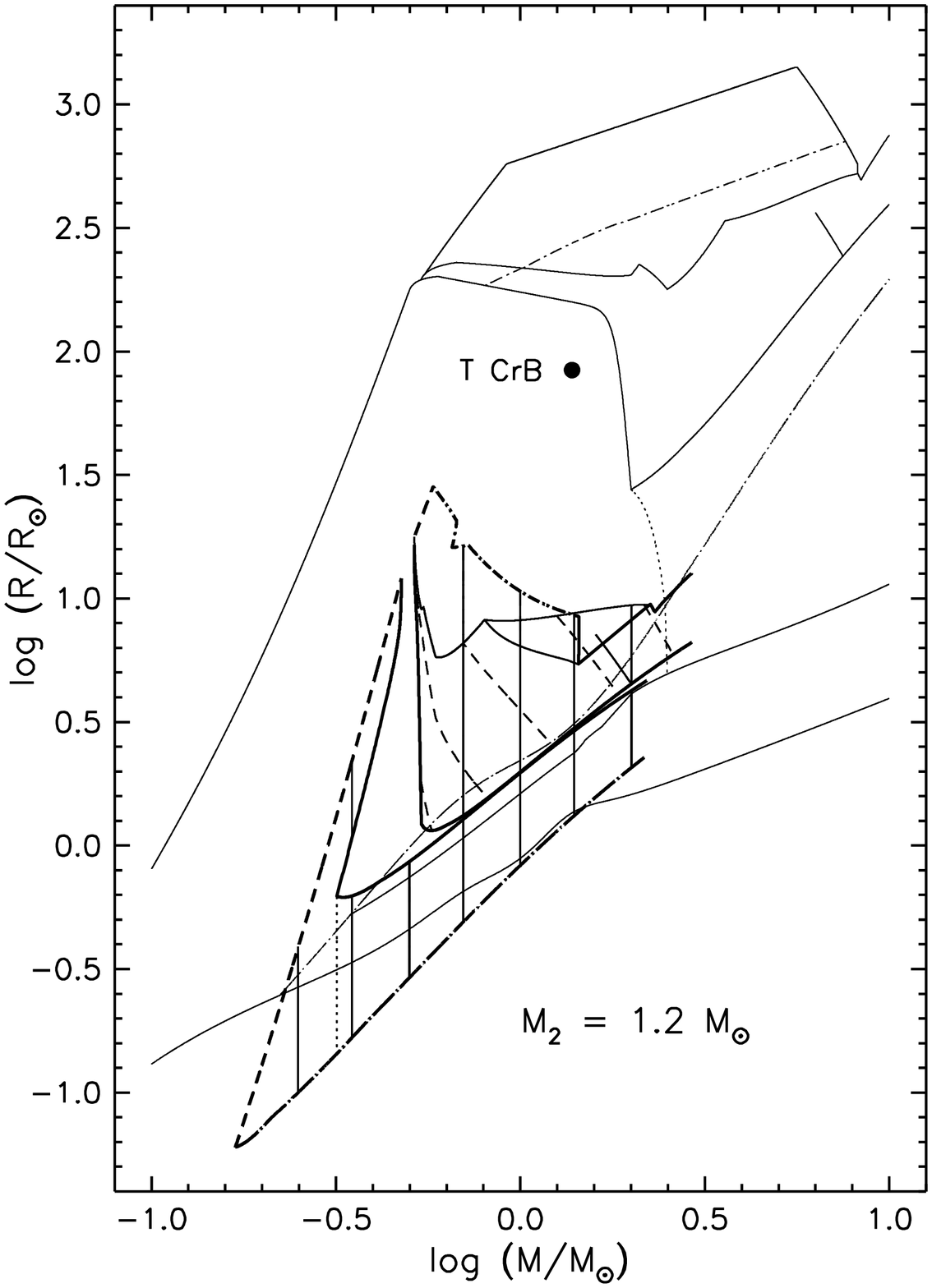}}
\caption{Post-common-envelope masses and Roche lobe radii for binaries
consisting of a white dwarf or helium star plus a $1.2\,M_{\odot}$ companion,
computed with $\alpha_{\rm CE} = 1$.  Remnant systems inhabit the regions
outlined in bold, and spanned vertically by lines of constant white
dwarf/helium star mass of 0.25, 0.35, 0.5, 0.7, 1.0, 1.4, and
$2.0\,M_{\odot}$.  Other initial sequences, encoded as in
Fig.~\ref{Mcore_diagram}, have been mapped through common envelope evolution. 
The location in this diagram of the white dwarf in the recurrent nova T CrB is
also indicated.}
\label{PostCEalpha}
\end{FPfigure}

It is evident that these long-period binaries are able to tap some energy
source not reflected in the energy budget in~(\ref{Afalpha}).  One
possibility, discussed repeatedly in studies of planetary nebula
ejection~(\cite{Luc67,Pac68},  more recently~\cite{Wag94,Han94,Han95}) is that
the recombination energy of the envelope comes into play.  For solar
composition material (and complete ionization), that recombination energy
amounts to $15.4\,{\rm eV\,amu^{-1}}$, or $1.49 \times 10^{13}\,{\rm
erg\,g^{-1}}$.  For tightly-bound envelopes on the initial giant branch of the
donor, this term is of little consequence; but near the tip of the low-mass
giant branch, and on the upper aymptotic giant branch of intermediate-mass
stars, it can become comparable with, or even exceed, the gravitational
potential energy of the envelope.  In the model calculations of
thermally-pulsing asymptotic giant branch stars by Wagenhuber \&
Weiss~\cite{Wag94}, the threshold for spontaneous ejection by envelope
recombination occurs consistently when the stellar surface gravity at the peak
thermal pulse luminosity falls to 
\begin{equation}
\label{grecomb}
\log g_{\rm HRI} = -1.118 \pm 0.042\ .
\end{equation}
This threshold marks the presumed upper limit to the radii of lower-mass
asymptotic giant branch stars in Figs.~\ref{MRdiagram} et seq. in the present
paper.  In fact, the total energies of the envelopes of these stars formally
becomes decidedly positive even before the onset of the superwind phase, also
shown in these figures.

Whether single stars successfully tap this ionization energy in ejecting
planetary nebulae is still debated, but the circumstances of mass transfer in
binary systems would seem to provide a favorable environment for doing so.  In
the envelopes of extended giants and asymptotic giant branch stars,
photospheric electron densities and opacities are dominated by heavy elements;
the middle of the hydrogen ionization zone is buried at optical depths of
order $\tau \sim 10^5$.  Adiabatic expansion of the envelope of the donor into
the Roche lobe of its companion can therefore trigger recombination even as
the recombination radiation is itself trapped and reprocessed within the flow,
much as the same process occurs in rising convective cells.

Other possible energy terms exist that have been neglected in the energetics
arguments above: rotational energy, tidal contributions, coulomb energy,
magnetic fields, etc.  But Virial arguments preclude most of these terms from
amounting to more than a minor fraction of the internal energy content of the
common envelope at the onset of mass transfer, when the energy budget is
established.  The only plausible energy source of significance is the input
from nuclear reactions.  In order for that input to be of consequence, it must
of course occur on a time scale short compared with the thermal time scale of
the common envelope.  Taam~\cite{Taa07} explored the possibility that shell
burning in an asymptotic giant branch core could be stimulated by mixing
induced dynamically in the common envelope (see also~\cite{Taa89,Taa94}). 
Nothing came of this hypothesis: mixing of fresh material into a burning shell
required taking low-density, high-entropy material from the common envelope
and mixing it downward many pressure scale heights through a strongly stable
entropy gradient to the high-density, low-entropy burning region.  In the face
of strong buoyancy forces, dynamical penetration is limited to scales of order
a pressure scale height.

\section{Common Envelope Evolution with Recombination}
\label{Recomb}

The notion that recombination energy my be of importance specifically to
common envelope evolution is not new.  It has been included, at least
parametrically in earlier studies, for example, by Han et al.~\cite{Han95},
who  introduced a second $\alpha$-parameter, $\alpha_{\rm th}$, characterizing
the fraction of the initial thermal energy content of the common envelope
available for its ejection.  The initial energy kinetic/thermal content of the
envelope is constrained by the Virial Theorem, however, and it is not clear
that there is a compelling reason for treating it differently from, say, the
orbital energy input from the inspiraling cores.  We choose below to formulate
common envelope evolution in terms of a single efficiency parameter, labeled
here $\beta_{\rm CE}$ to avoid confusion with $\alpha_{\rm CE}$ as defined
above.

By combining the standard stellar structure equations for hydrostatic
equilibrium and mass conservation, we can obtain an expression for the
gravitational potential energy, $\varOmega_{\rm e}$, of the common envelope:
\begin{equation}
\label{VirialTheorem}
\varOmega_{\rm e} \equiv - \int_{M_{\rm c}}^{M_{\ast}} \frac{GM}{r} \, \D M =
3 \left. PV \rule[-3pt]{0pt}{16pt} \right|_{R_{\rm c}}^{R_{\ast}} - 3
\int_{V_{\rm c}}^{V_{\ast}} P \, \D V ~ ,
\end{equation}
where subscripts $c$ refer to the core-envelope boundary, and $\ast$ to the
stellar surface.  This is, of course, the familiar Virial Theorem applied to a
stellar interior.

It is convenient to split the pressure in this integral into non-relativistic
(particle), $P_{\rm g}$, and relativistic (photon), $P_{\rm r}$, parts.  The
envelopes of giants undergoing common envelope evolution are sufficiently cool
and non-degenerate to make the classical ideal gas approximation an excellent
one for the particle gas.  One can then write
\begin{equation}
\label{pressure}
P = P_{\rm g} + P_{\rm r} = \frac{2}{3} u_{\rm g} + \frac{1}{3} u_{\rm r} \ ,
\end{equation}
where $u_{\rm g}$ and $u_{\rm r}$ are kinetic energy densities of particle and
radiation gases, respectively.  The \emph{total} internal energy density of
the gas is
\begin{equation}
\label{internalenergy}
u = u_{\rm g} + u_{\rm r} + u_{\rm int} \ ,
\end{equation}
where the term $u_{\rm int}$ now appearing represents non-kinetic
contributions to the total energy density of the gas, principally the
dissociation and ionization energies plus internal excitation energies of
bound atoms and molecules.  The overwhelmingly dominant terms in $u_{\rm int}$
are the ionization energies: $u_{\rm int} \approx \rho\chi_{\rm eff}$.

Integrating over the stellar envelope, we obtain for the total energy $E_{\rm
e}$ of the envelope:
\begin{eqnarray}
\label{EVirial}
E_{\rm e} & = & \varOmega_{\rm e} + U_{\rm e} \nonumber \\
& = & \left( 3 \left. P U \right|_{R_{\rm c}}^{R_{\ast}} - 2 U_{\rm g} -
U_{\rm r} \right) + \left( U_{\rm g} + U_{\rm r} + U_{\rm int} \right)
\nonumber \\
& = & - 4 \pi R_{\rm c}^3 P_{\rm c} - U_{\rm g} + U_{\rm int} \ ,
\end{eqnarray}
where we explicitly take $P_{\ast} \rightarrow 0$.  In fact, experience shows
that, for red-giant like structures, $R_{\rm c}$ is so small that the first
right-hand term in the last equality can generally be neglected.  In that
case, we get the familiar Virial result, but with the addition of a term
involving the ionization/excitation/dissociation energy available in the gas,
$U_{\rm int} \approx M_{\rm e} \chi_{\rm eff}$, which becomes important for
diffuse, loosely-bound envelopes.

\begin{FPfigure}
\setlength\fboxsep{0pt}
\centering
\resizebox{.97\textwidth}{!}{\includegraphics{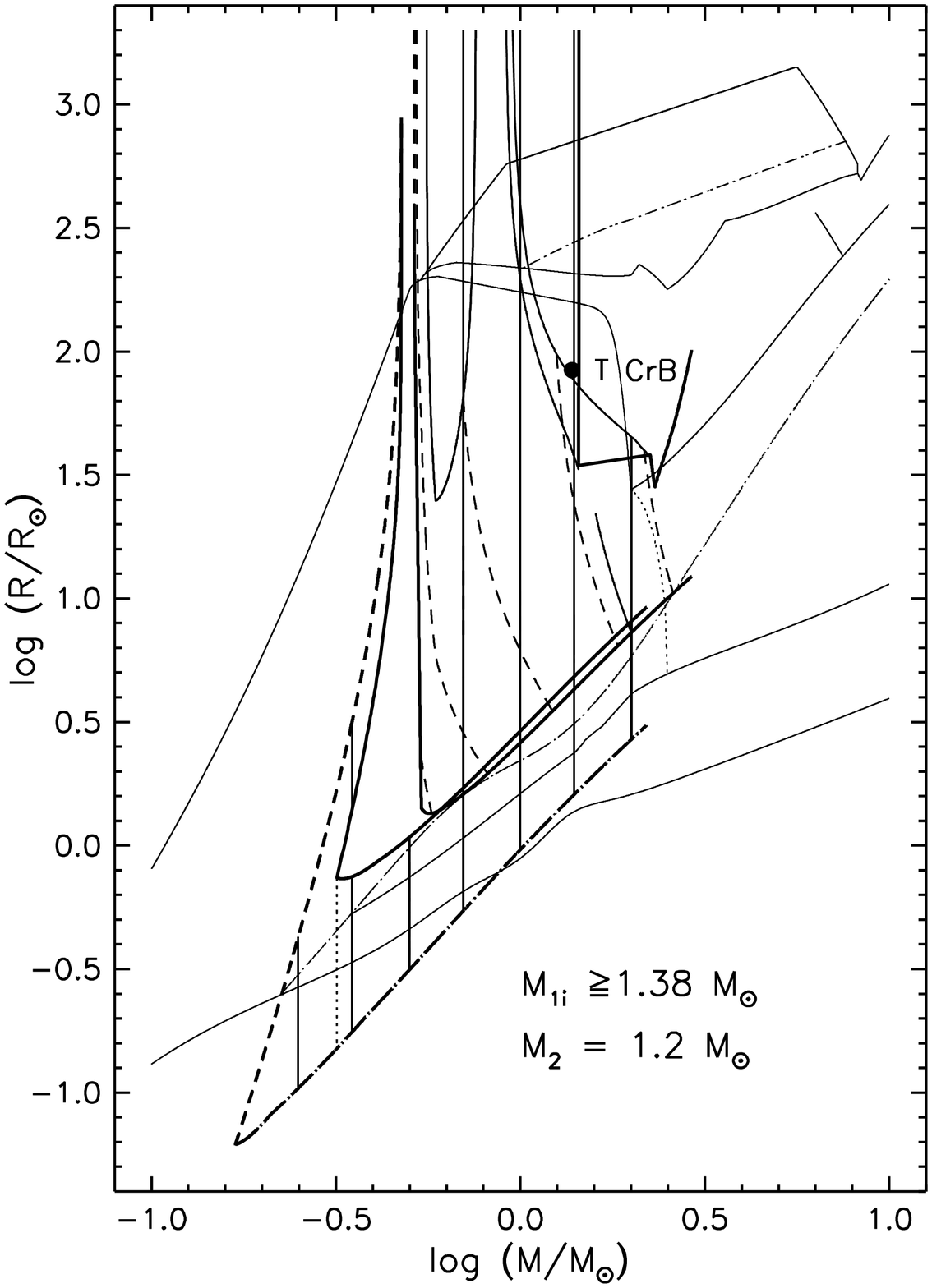}}
\caption{Post-common-envelope masses and Roche lobe radii as in
Fig.~\ref{PostCEalpha}, but with recombination energy included, computed from
(\ref{Afbeta}) with $\beta_{\rm CE} = 1$, with the approximation
$2\lambda_{\varOmega} = \lambda_{\rm P} = \lambda$ from (\ref{lambda}).  At
small separations, the differences are inconsequential, but substantially
larger final separations are allowed when $A_{\rm f} \gtrsim 10\,R_{\odot}$
($R_{\rm L} \gtrsim 3\,R_{\odot}$).}
\label{PostCEbeta}
\end{FPfigure}

In the context of common envelope evolution, it is of course the dissipated
orbital energy, $E_{\rm orb}^{\rm (i)} - E_{\rm orb}^{\rm (f)}$, that must
unbind the envelope.  However, the inclusion of $U_{\rm int}$ in $E_{\rm e}$
now opens the possibility that the common envelope began with \emph{positive}
total energy; that is, in the usual $\alpha_{\rm CE}$-prescription, it is
possible for $\lambda^{-1}$ to be zero or even negative, which has the
undesirable consequence that $\alpha_{\rm CE}$ need not lie in the interval $0
\le \alpha_{\rm CE} \le 1$ for all physically-possible outcomes.  However, the
gravitational potential energy of the envelope, $\varOmega_{\rm e}$, is
negative-definite, and by comparing it with all available energy sources
(orbital energy released plus internal energy of the envelope), we can define
an ejection efficiency $\beta_{\rm CE}$ that has the desired property, $0 \le
\beta_{\rm CE} \le 1$:
\begin{equation}
\label{betaCE}
\beta_{\rm CE} \equiv \frac{\varOmega_{\rm e}}{(E_{\rm orb}^{\rm (f)} - E_{\rm
orb}^{\rm (i)}) - U_{\rm e}} = \frac{4\pi R_{\rm c}^3 P_{\rm c} + 2 U_{\rm g} 
+ U_{\rm r}}{E_{\rm orb}^{\rm (i)} - E_{\rm orb}^{\rm (f)} + U_{\rm g} +
U_{\rm r} + U_{\rm int}} \ .
\end{equation}
By analogy to the form factor $\lambda$ in the conventional $\alpha_{\rm CE}$
formalism above, we can define separate form factors $\lambda_{\varOmega}$ for
the gravitational potential energy and $\lambda_{\rm P}$ for the gas plus
radiation contributions to the (kinetic) internal energy of the envelope:
\begin{equation}
\label{Omegaenv}
\varOmega_{\rm e} = \frac{GMM_{\rm e}}{\lambda_{\varOmega} R} \ \ \ \ \
\mbox{and} \ \ \ \ \ U_{\rm g} + U_{\rm r} = \frac{GMM_{\rm e}}{\lambda_{\rm
P} R} \ ,
\end{equation}
In contrast, the recombination energy available can be written simply in terms
of an average ionization energy per unit mass, 
\begin{equation}
\label{Uint}
U_{\rm int} = M_e \chi_{\rm eff} \ .
\end{equation}
The ratio of final to initial orbital separation then becomes
\begin{equation}
\label{Afbeta}
\frac{A_{\rm f}}{A_{\rm i}} = \frac{M_{\rm 1c}}{M_1} \left[ 1 + 2 \left(
\frac{1}{\beta_{\rm CE} \lambda_{\varOmega} r_{\rm 1,L}} -
\frac{1}{\lambda_{\rm P} r_{\rm 1,L}} - \frac{\chi_{\rm eff} A_{\rm i}}{GM_1}
\right) \left( \frac{M_1 - M_{\rm 1c}}{M_2} \right) \right]^{-1} \ .
\end{equation}
In the limit that radiation pressure $P_{\rm r}$, ionization energy ($U_{\rm
int}$), and the boundary term ($4\pi R_{\rm c}^3 P_{\rm c}$) are all
negligible, then $2 \lambda_{\varOmega} \rightarrow \lambda_{\rm P}
\rightarrow \lambda$ and $\beta_{\rm CE} \rightarrow 2 \alpha_{\rm CE}/(1 +
\alpha_{\rm CE})$.

The ability to tap the recombination energy of the envelope has a profound
effect on the the final states of the longest-period intermediate-mass
binaries, those that enter common envelope evolution with relatively massive,
degenerate carbon-oxygen (or oxygen-neon-magnesium) cores.  As is evident in
Fig.~\ref{PostCEbeta}, possible final states span a much broader range of
final orbital separations.  Indeed, for the widest progenitor systems, the
(positive) total energy of the common envelope can exceed the (negative)
orbital energy of the binary, making arbitrarily large final semimajor axes
energetically possible.\footnote{The final orbit remains constrained by the
finite initial orbital angular momentum of the binary.  Final semimajor axes
much in excess of the initial semimajor axis may be energetically allowed, but
the finite angular momentum available means that they cannot be circular --
see~(\ref{JfJi}) -- an effect which has been neglected in
Fig.~\ref{PostCEbeta}.}  The inclusion of recombination energy brings both
T~CrB and RS~Oph within energetically accessible post-common-envelope states. 
It suffices as well to account for the exceptionally long-period close double
white binary PG~1115+166, as suggested by Maxted, et al.~\cite{Max02}.

\section{Conclusions}
\label{Conclusions}

Re-examination of global constraints on common envelope evolution leads to the
following conclusions:

Both energy and angular momentum conservation pose strict limits on the
outcome of common envelope evolution.  Of these two constraints, however,
energy conservation is much the more demanding.

The recent study of close double white dwarf formation by Nelemans \&
Tout~\cite{Nel05} shows clearly that their progenitors can have lost little
orbital energy through their first episodes of mass transfer.  Since common
envelope ejection must be rapid if it is to be efficient, its energy budget is
essentially fixed at its onset by available thermal and gravitational terms. 
The preservation of orbital energy through that first phase of mass transfer
therefore indicates that the observed close double white dwarfs escaped common
envelope formation in that first mass transfer phase.  They evidently evolved
through quasi-conservative mass transfer.  However, strictly mass- and angular
momentum-conservative mass transfer leaves remnant accretors that are too
massive and compact to account for any but the shortest-period close double
white dwarfs.  Significant mass loss and the input of orbital energy prior to
the onset of the second (common envelope) phase of mass transfer are required. 
The requisite energy source must be of nuclear evolution, which is capable of
driving orbital expansion and stellar wind losses during the slower (thermal
recovery or nuclear time scale) phases of quasi-conservative mass transfer, or
during the interval between first and second episodes of mass transfer. 
Details of this process remain obscure, however.

Long-period cataclysmic variables such as T~CrB and RS~Oph pose a more extreme
test of common envelope energetics.  With their massive white dwarfs, the
evident remnants of much more massive initial primaries, they are nevertheless
too low in total systemic mass to be plausible products of quasi-conservative
mass transfer, but too short in orbital period to have escaped tidal mass
transfer altogether.  They must be products of common envelope evolution, but
to have survived at their large separations, they demand the existence of a
latent energy reservoir in addition to orbital energy to assist in envelope
ejection.  It appears that these binaries efficiently tap
ionization/recombination energy in ejecting their common envelopes.  That
reservoir is demonstrably adequate to account for the survival of these
binaries.  Its inclusion requires only a simple revision to the
parameterization of common envelope ejection efficiency.

\begin{acknowledgement}
This work owes its existence to both the encouragement and the patience of
Gene Milone, to whom I am most grateful.  Thanks go as well to Ron Taam for a
useful discussion of possible loopholes in common envelope theory, and to
Jarrod Hurley for providing the source code described in Hurley, et al.
(2000).  This work was supported in part by grant AST 0406726 to the
University of Illinois, Urbana-Champaign, from the US National Science
Foundation.
\end{acknowledgement}

\end{document}